\documentclass[twoside]{article}
\usepackage{amsmath}
\usepackage[psamsfonts]{amssymb}
\usepackage{cmmib57}
\newcommand{\bPf}{\par\vspace*{-4pt}\indent{\sc Proof.}\enskip}
\newcommand{\ePf}{\medskip}
\def\QED{\hskip0.1em\hfill\null\ \null\nobreak\hfill\kern3pt\vbox{\hrule\hbox
   {\vrule\kern1pt\vbox{\kern1.7pt\hbox{$\scriptscriptstyle{QED}$}
    \kern0.2pt}\kern1pt\vrule}\hrule}}

\def\END{\hskip0.1em\hfill\null\ \null\nobreak\hfill\kern3pt\vbox{\hrule\hbox
   {\vrule\kern1pt\vbox{\kern1.7pt\hbox{$\,\,\,\vspace{5pt}$}
    \kern0.2pt}\kern1pt\vrule}\hrule}}
\newtheorem{theorem}{Theorem}
\newtheorem{lemma}{Lemma}
\newtheorem{corollary}{Corollary}
\newtheorem{proposition}{Proposition}
\newtheorem{remark}{Remark}
\newtheorem{definition}{Definition}
\newtheorem{example}{Example}
\newcommand{\bCd}{\bEq\begin{CD}}
\newcommand{\eCd}{\end{CD}\eEq}
\newcommand{\bcd}{\beq\begin{CD}}
\newcommand{\ecd}{\end{CD}\eeq}
\newcommand{\ben}{\begin{enumerate}}
\newcommand{\een}{\end{enumerate}}
\newcommand{\bEq}{\begin{eqnarray}}
\newcommand{\eEq}{\end{eqnarray}}
\newcommand{\beq}{\begin{eqnarray*}}
\newcommand{\eeq}{\end{eqnarray*}}
\newcommand{\bDf}{\begin{definition}\em}
\newcommand{\eDf}{\end{definition}}
\newcommand{\bLm}{\begin{lemma}}
\newcommand{\eLm}{\end{lemma}}
\newcommand{\bPr}{\begin{proposition}}
\newcommand{\ePr}{\end{proposition}}
\newcommand{\bTh}{\begin{theorem}}
\newcommand{\eTh}{\end{theorem}}
\newcommand{\bCr}{\begin{corollary}}
\newcommand{\eCr}{\end{corollary}}
\newcommand{\bRm}{\begin{remark}\em}
\newcommand{\eRm}{\end{remark}}
\newcommand{\bEx}{\begin{example}\em}
\newcommand{\eEx}{\end{example}}

\newcommand{\Z}{\mathbb{Z}}
\newcommand{\ie}{{\em i.e$.$} }
\newcommand{\eg}{{\em e.g$.$} }
\newcommand{\R}{I\!\!R}
\newcommand{\A}{\forall}

\DeclareMathOperator{\byd}{{\raisebox{.1ex}{:}{=}}}

\newcommand{\bbet}{\boldsymbol{\bet}}

\newcommand{\bxi}{\boldsymbol{\xi}}

\newcommand{\cA}{\mathcal{A}}
\newcommand{\cB}{\mathcal{B}}
\newcommand{\cC}{\mathcal{C}}
\newcommand{\cD}{\mathcal{D}}

\newcommand{\cH}{\mathcal{H}}

\newcommand{\cK}{\mathcal{K}}
\newcommand{\cL}{\mathcal{L}}
\newcommand{\cM}{\mathcal{M}}
\newcommand{\cN}{\mathcal{N}}

\newcommand{\cP}{\mathcal{P}}

\newcommand{\bu}{\boldsymbol{u}}

\newcommand{\bz}{\boldsymbol{z}}

\newcommand{\bE}{\boldsymbol{E}}
\newcommand{\bF}{\boldsymbol{F}}
\newcommand{\bG}{\boldsymbol{G}}

\newcommand{\bK}{\boldsymbol{K}}
\newcommand{\bL}{\boldsymbol{L}}

\newcommand{\bP}{\boldsymbol{P}}

\newcommand{\bU}{\boldsymbol{U}}
\newcommand{\bV}{\boldsymbol{V}}

\newcommand{\bZ}{\boldsymbol{Z}}

\newcommand{\bbeta}{\boldsymbol{\beta}}

\newcommand{\bet}{\beta}

\newcommand{\del}{\delta}

\newcommand{\ome}{\omega}

\newcommand{\bPsi}{\boldsymbol{\Psi}}
\DeclareMathOperator{\T}{T}

\title{{\bf Constructing towers with skeletons from open Lie algebras and integrability
}}
\author{{\normalsize 
M. Palese and E. Winterroth}
\\{\footnotesize Department of Mathematics,
University of Torino}
\\{\footnotesize Via C. Alberto 10, 10123 Torino, Italy}\\ 
{\footnotesize e--mails: 
{\sc 
[marcella.palese, ekkehart.winterroth]@unito.it}}}
\date{}
\begin{document}

\maketitle

\begin{abstract}
We provide a given algebraic structure with the structure of an infinitesimal algebraic skeleton.
The necessary conditions for  integrability of the absolute parallelism of a tower with such a skeleton  are dispersive nonlinear models and related conservation laws given in the form of associated linear spectral problems.

\medskip


\end{abstract}

\section{Introduction}

It is nowadays well recognized that {\em algebraic properties} of nonlinear systems, which  play a role in a variety of physical phenomena, are relevant from the point of view of integrability. This is a topic which is far from being trivial for both discrete and continuous, as well as, classical and quantistic models.
The origin of this relevance lies in the concept of integrability as of having `enough' conservation laws to exaustively describe the dynamics and from a variational point of view can be seen as a version of  the inverse Noether Theorem.
Historically, in fact, the algebraic-geometric  approach is based on the {\em request of the 
existence of conservation laws} which leads to the existence of symmetries (in terms of algebraic structures).

In this paper we will consider such algebraic structures also called `open' Lie algebra structures, in the sense that not all the commutators (\ie not all the Lie algebra structure constants) are determined.
Algebraic structures of such type appear constructively 
as an outcome of a geometric approach, based on the concept of Cartan's moving frames, proposed by Wahlquist and Estabrook 
for the study of integrability properties of nonlinear dispersive systems. They were specifically related with the existence of an infinite set of associated conservation laws generated by pseudopotentials \cite{WaEs75,EsWa76}. In their approach, conservation laws  are written in terms of `prolongation' forms and the algebraic outcome was called a `prolongation structure'. The latter appeared immediately  very rich and -
by using their words - was shown to `contain' in particular  linear inverse scattering equations 
and B\"acklund transformations (see in particular \cite{WaEs73}). 
By using recurrence relations, they conjectured that the structure was `open', that is not a set of structure relations of a finite--dimensional Lie group, as they wrote. Since then, `open' Lie algebras have been extensively studied in order to distinguish them from freely generated infinite-dimensional Lie algebras.  

Such nonlinear prolongation algebras can be thus related to integrable nonlinear 
field equations which are geometrically expressed by means of closed differential ideals and they arise via the introduction of an arbitrary number of 
`prolongation' forms containing pseudopotentials as new dependent variables, and by requiring the algebraic equivalence between the generators of the `prolonged' ideal and its exterior differential, \ie by requiring an
integrability condition for the prolonged differential ideal.
The geometric interpretation of pseudopotentials and prolongation structures was the object of various studies, among them the seminal papers \cite{He76,EsWaHe77,PRS77,PRS79}. Attempting a {\em description of symmetries in terms of Lie algebras} implies the appearing of an homogeneous space and thus the interpretation of prolongation forms as {\em Cartan--Ehresmann connections} \cite{Eh51}; in \cite{Ryb02} the concept of a {\em special connection} was pointed out; see also \cite{PaWi02,PaWi03,Pa05}. 

Within the theory of moving frames, Estabrook also investigated the duality between vector fields and forms, and applied it to the algebraic structure obtained by the prolongation technique \cite{Es82}. By reformulating geometrically  the prolongation procedure in terms of a connection on a fiber bundle, an incomplete Lie algebra of vector fields was derived. 
Conversely, given an `open' algebra, he showed that one can derive a corresponding differential ideal (actually a whole `family' of differential ideals among which also the original one, from  which the open structure itself could be derived as its prolongation structure). 
The latter aspect has been investigated subsequently by many authors; see \eg  \cite{LeSo89,Pa93,PaWi02} and references therein.

It should be stressed that within this approach the `unknowns' are {\em both} conservation laws and symmetries and it is clear that the main point in this is how to realize the form of conservation laws 
and thus the explicit expression of prolongation forms.
Different formulations of the prolongation ideal lead to both different algebraic structures (symmetries) and corresponding conservation laws: of course, also
depending on the postulated structure of prolongation forms, we could obtain `open' or `closed' Lie algebras. 

We use the algebraic properties of a `laboratory' model to explicate 
an algebraic-geometric interpretation of the above mentioned `prolongation' procedure 
in terms of towers with infinitesimal algebraic skeletons \cite{Mo93}.
We shall consider a given abstract open Lie algebra structure, we show that it can be caracterized as an infinitesimal algebraic skeleton on an appropriate vector space and construct a tower with such a skeleton and the functorially associated gauge-natural bundle, the sections of which can be used to pull-back exterior differential forms to nonlinear dispersive field equations.  

In particular, by realizing infinite dimensional open Lie algebra structures as infinitesimal algebraic skeletons we obtain a family of coupled nonlinear Schroedinger equations describing several nonlinear phenomena, in particular, models of waves propagation in optical fibers and corresponding conservation laws in the form of a spectral linear problem associated with them is explicitly written \cite{AbSe81}. 

It is noteworthy that we obtain as an element of the family the well known Manakov equation which then has in common with coupled nonlinear Schroedinger equations {\em the same tower with skeleton}. This fact is much stronger than knowing that solutions of the coupled nonlinear Schroedinger equations could be reconstructed (resorting to transformations) by solutions of the Manakov equation. 

\section{An infinitesimal algebraic skeleton from an open Lie algebra structure}

As mentioned above,  we intend to explicate 
an algebraic-geometric interpretation of the `prolongation' procedure 
in terms of towers with infinitesimal algebraic skeletons.

It is well known that the notion of an (infinitesimal) algebraic skeleton is an abstraction
of some algebraic aspects of the homogeneous spaces \cite{Mo93}.
As a prototype of skeleton, one can consider a homogeneous space $\bL/\bK$,
where $\bL$ is a finite-dimensional Lie group with Lie algebra $\mathfrak{l}$
and $\bK$ is a closed subgroup of $\bL$ with Lie algebra $\mathfrak{k}$. The
triple $(\mathfrak{l},\bK, Ad)$ becomes a skeleton on
$\bV=\mathfrak{l}/\mathfrak{k}$ by identifying
$\mathfrak{l}=\bV\oplus\mathfrak{k}$.

Let then $\bV$ denotes a finite--dimensional vector space. 
An {\em algebraic skeleton} on $\bV$ is a triple $(\bE,\bG,\rho)$, with 
$\bG$ a
(possibly infinite-dimensional) Lie group, 
$\bE=\bV\oplus\mathfrak{g}$, 
$\mathfrak{g}$ the Lie
algebra of $\bG$, and 
$\rho$ a representation of $\bG$ on $\bE$ (infinitesimally of $\mathfrak{g}$ on $\bE$) such that
$\rho(g)x=Ad(g)x$, for $g\in\bG$, $x\in\mathfrak{g}$. The generalization involved in the notion of a skeleton is that, $\bG$ can be infinite dimensional and that $\bE$ is not necessarily equipped with a Lie algebra structure. 

Let  $\bZ$ be a manifold of type
$\bV$ (\ie  $\A \bz\in \bZ$,  $T_{\bz} \bZ \simeq \bV$).
We say that a  principal fibre bundle
$\bP(\bZ,\bG)$ provided with an 
{\em  absolute parallelism} $\omega$ on $\bP$
is a {\em tower} on $\bZ$ with skeleton $(\bE, \bG,\rho)$
 if $\omega$ takes values in $\bE$
and satisfies: 
 $R^{*}_{g}\omega = \rho(g)^{-1}\ome$, for
$g\in \bG$; $\ome(\tilde{A}) = A$, for $A\in\mathfrak{g}$;
here $R_{g}$ denotes the right translation and $\tilde{A}$ the fundamental
vector field induced on $\bP$ from $A$.
In general, the absolute parallelism  {\em does not}  define a Lie algebra homomorphism.

Let $\mathfrak{l}$ be a Lie algebra and $\mathfrak{k}$ a Lie subalgebra of
$\mathfrak{l}$. Let $\bK$ be a Lie group with Lie algebra $\mathfrak{k}$ and $\bP(\bZ, \bK)$ be a principal fibre bundle with structure group $\bK$ over a manifold $\bZ$ as above. 
A {\em Cartan connection} in $\bP$ of type
$(\mathfrak{l}, \bK)$ is 
a $1$--form $\omega$ on $\bP$ with values in
$\mathfrak{l}$ satisfying the following conditions: 
- $\omega |_{T_{u} \bP}: T_{u} \bP\to \mathfrak{l}$ is an isomorphism $\forall u\in
\bP$; 
- $R^{*}_{g}\omega=Ad(g)^{-1}\omega$ for $g\in \bK$;
- $\omega(\tilde{A})=A$ for $A \in \mathfrak{k}$.
A  Cartan connection
$(\bP, \bZ, \bK, \omega)$ of type $(\mathfrak{l}, \bK)$ is a special case of a tower on $\bZ$.

Remark that since, {\em a priori}, the prolongation algebra does not close into a Lie algebra the starting point (\ie the construction of prolongation forms) for the prolongation procedure should be considered simply only a tower (with an absolute parallelism) and in general not, at least a {\em a priori}, a Cartan connection.
Thus, in principle, we should  think of Estabrook-Wahlquist's  prolongation forms as {\em absolute parallelism forms}.
The corresponding open Lie algebra  structure can be provided of the 
structure of an infinitesimal algebraic skeleton on a suitable space.
First we have to prove that a finite dimensional space $\bV$ and a Lie algebra $\mathfrak{g}$ exist satisfying the definition of a skeleton, \ie in particular that a suitable representation $\rho$ can be defined. 
The representation $\rho$ (\ie the way in which the Lie algebra $\mathfrak{g}$ acts on the vector space $\bV$) is obtained by means of the request of integrability for the {\em absolute parallelism} of a tower on $\bZ$, with skeleton  $(\bE, \bV, \mathfrak{g})$. 

Let $\kappa,\mu$  be real parameters,
$\sigma= ||\sigma_{lm}||$, a matrix with $\sigma_{lm}=0$, for $l=m$ and $\sigma_{lm}=1$, for $l \neq m$. 
Let us consider the following algebraic structure which, for reasons which will be clear later, we just denote by $\bE$:
\beq
& [\chi_{1},\chi_{2}]=0 \,,  [\chi_{3},\chi_{4}]=0 \,,   [\chi_{5},\chi_{6}]=0\,, [ [\chi_{4},\chi_{1}],\chi_{4}]= 0 \,, 
\\ 
& [ [\chi_{1},\chi_{5}],\chi_{1}]= 0\,, [ [\chi_{1},\chi_{5}],\chi_{2}]= 0 \,,  [ [\chi_{2},\chi_{5}],\chi_{1}]= 0\,, 
\\
& [ [\chi_{2},\chi_{5}],\chi_{2}]= 0 \,,  [ [\chi_{3},\chi_{5}],\chi_{4}]= 0\,, [ [\chi_{4},\chi_{5}],\chi_{3}]= 0 \,, 
\\
& [ [\chi_{3},\chi_{5}],\chi_{3}]= 0\,, [ [\chi_{4},\chi_{5}],\chi_{4}]= 0 \,,  [ [\chi_{3},\chi_{1}],\chi_{5}]= 0\,, 
\\
&[ [\chi_{4},\chi_{1}],\chi_{5}]= 0 \,,  [ [\chi_{3},\chi_{2}],\chi_{5}]= 0\,, [ [\chi_{4},\chi_{2}],\chi_{5}]= 0 \,, 
\\
& [ [\chi_{3},\chi_{2}],\chi_{2}]= 0\,, [ [\chi_{4},\chi_{1}],\chi_{1}]= 0 \,, [ [\chi_{3},\chi_{2}],\chi_{3}]= 0\,,  
\\
& [ [\chi_{3},\chi_{1}],\chi_{1}]= \mu \chi_{1}\,, [ [\chi_{4},\chi_{1}],\chi_{2}] = \frac{1}{2}\mu\chi_{2}  \,, 
\\
& [ [\chi_{3},\chi_{2}],\chi_{1}] = \frac{1}{2}\mu\chi_{1} \,,  [ [\chi_{4},\chi_{2}],\chi_{2}]= \mu \chi_{2}\, 
\\
&  [ [\chi_{3},\chi_{1}],\chi_{3}]= - \mu \chi_{3}  \,,  [ [\chi_{3},\chi_{1}],\chi_{4}]= - \frac{1}{2} \mu \chi_{3} \,,
\\
& [ [\chi_{3},\chi_{2}],\chi_{4}]= - \frac{1}{2} \mu \chi_{4}\,,  [ [\chi_{4},\chi_{3}],\chi_{4}]= - \frac{1}{2} \mu \chi_{4} \,, 
\\
&  [ [\chi_{4},\chi_{1}],\chi_{3}]= - \frac{1}{2} \mu \chi_{3} \,,  [ [\chi_{4},\chi_{2}],\chi_{4}]= - \mu \chi_{4}  \,,
\\
& i \kappa \sigma \chi_{j} + i [ [\chi_{j},\chi_{5}],\chi_{5}] - [\chi_{6},\chi_{j}] = 0 \,, j=1,2,3,4\,. 
\eeq
It is a (possibly infinite dimensional) vector space  and in particular it has the structure of an open Lie algebra. 
We show how it  can be provided with the structure of an infinitesimal algebraic skeleton on a finite dimensional vector space $\bV$. The following Lemma provides what is also sometimes called an infinite-dimensional `realization' of $\bE$.

\bLm
There exists a homomorphis $\cH$ between $\bE$ and an infinite-dimensional Lie algebra with a loop structure of the Kac-Moody type which we denote by $\mathfrak{g}$:
\beq
\left[ \T_{lm}^{(n)},\T_{\bar{l} \bar{m}}^{(\bar{n})}\right]\,=\,
i \del_{\bar{m} l}\T_{\bar{l} m}^{(n+\bar{n})}\,-\,i
\del_{m\bar{l}}\T_{l\bar{m}}^{(n+\bar{n})}\,.
\eeq
\eLm
\bPf
The homomorphism $\cH$ is defined by the following relations
\beq
& \chi_{k} \rightarrow  \T_{0k}^{(n)}\,,  \chi_{m} \rightarrow \T_{m0}^{(-n)} \,,
 t_{lm} \rightarrow i \T_{lm}^{[n(\del_{l0} - \del_{m0})]}\,,
 \chi_{5} \rightarrow \lambda\left(\T_{11}^{(l)} + \T_{22}^{(l)}
\right)\,,
\\
& \chi_{6} \rightarrow -3i \lambda^{2}\left(\T_{11}^{(2l)} 
+ \T_{22}^{(2l)}\right) -i \kappa\left(\T_{12}^{(0)} + 
\T_{21}^{(0)}\right)\,,
\eeq
where $k=1,2$, $m=3,4$, $n,l \in \Z$, $\lambda$ is a parameter and we put $t_{lm}= [\chi_{l},\chi_{m}]$.

If, in particular, $l,m,\bar{l}, \bar{m}\,\ne\,0$, the vector fields $\T_{lm}^{(n)}$ satisfy the commutation relations defining $\mathfrak{g}$.
\ePf

\bDf
We define the vector space $\bV$ to be  the kernel of the above homomorphism, \ie 
$\bV\byd\textstyle{ker}\cH$.
\eDf

Let us consider a manifold $\bP$ on which the Lie group $\bG$, with  Lie algebra $\mathfrak{g}$, acts on the right; $\bP$ is a principal bundle $\bP\to\bZ\simeq\bP/\bG$. By construction, we have that $\bZ$ is a manifold of type
$\bV$.
A tower $\bP(\bZ,\bG)$ on $\bZ$
 with skeleton $(\bE, \bG,\rho)$ is an 
{\em  absolute parallelism} $\omega$ on $\bP$ valued in $\bE$. 
Suppose a left action of $\bG$ on a manifold $\bU$ of type $\bE$ (\ie $T_{\bu}\bU\simeq\bE$) which induces a right action on the product manifold $J_{1}\bP\times \bU \to \bZ $. We have a functorial construction of the associated gauge-natural bundle 
$J_{1}\bP\times_{\bG}\bU \to \bZ$.
Let us now consider the induced absolute parallelism given by forms locally exspessed as (a superscript is understood, its range  being the dimension of the vector space $\bE$, while $(x,t,\bbeta,\bbeta^{*})$ are local coordinates in $\bU$):
\beq
\omega =d\xi+{\bF}(\bbeta,\bbeta^{*},\bbeta_{t},\bbeta^{*}_{t},\bbeta_{x},\bbeta^{*}_{x}; \xi)d x
+{\bG}(\bbeta,\bbeta^{*},\bbeta_{t},\bbeta^{*}_{t},\bbeta_{x},\bbeta^{*}_{x}; \xi)d t \,,.
\eeq

The important point is now that the way how the functions $\bF$ and $\bG$ are determined specifies how the representation $\rho$ of our skeleton is defined. In fact, we cannot  speak  of $\omega$ as the absolute parallelism for a tower unless we define the skeleton over which the tower is constructed, \ie unless we define the skeleton representation $\rho$. 

The absolute parallelism  is a connection on $\bP$ valued in $\bE$ and invariant under the right action of $\bG$ which implies  that there is a right action of $\bG$ on $\bV$.
Keeping in mind the request of invariance (which implies a more specific form of $\bF$ and $\bG$ as a summation of product of functions of $\bbeta,\bbeta^{*}$ and their derivatives, from one side and functions of $\xi$ on the other side), different {\em ans\"atze} can be made on ${\bF}$ and  ${\bG}$.
A choice can be, for example, $\bF = - i \sum^{2}_{k=1} [\cA_{k} \T^{(n)}_{0k} + \cB_{k} \T^{(-n)}_{k0} $ $+$ $ \cC_{k} \T^{(n+l)}_{0k}$ $+$ $\cD_{k} \T^{(-n+l)}_{k0} $ $+$ $ \cP_{k} \T^{(0)}_{1k} $ $ + $ $ \cK_{k} \T^{(0)}_{2k} $ $ + $ $ 3i \T^{(2l)}_{kk}]$, 
$\bG = \sum^{2}_{k=1} [\beta_{k} \T^{(n)}_{0k} $ $+$ $\epsilon \beta^{*}_{k} \T^{(-n)}_{k0} 
$ $-$ $i \T^{(l)}_{kk}]$,
where $\epsilon$ is a real quantity;  $\cA_{k}$, $\cB_{k}$, $\cC_{k}$, $\cD_{k}$, $\cP_{k}$, $\cK_{k}$ 
are functions of $\beta_{j}$, $\beta^{*}_{j}$ to be determined.

\bPr
The representation $\rho$ is defined by requiring that $\cA_{k}$, $\cB_{k}$, $\cC_{k}$, $\cD_{k}$, $\cP_{k}$, $\cK_{k}$ 
are functions of $\beta_{j}$, $\beta^{*}_{j}$ in such a way that 
the operators $\T (\cdot)$ satisfy the Ka\v{c}-Moody algebra 
so that $\omega=0$ along the sections $\beta_{j} (t, x) $ and $\beta_{j}^{*} (t, x) $ and that the 
integrability condition $\xi^{(i)}_{xt}\,=\,\xi^{(i)}_{tx}$ 
holds true.
\ePr
\bPf
This integrability condition for the absolute parallelism defines the vector space $\bV$, and more precisely  the decomposition $\bE=\bV\oplus\mathfrak{g}$ in such a way  that $\bV$ is isomorphic to the horizontal subspace which is the lift of $T_{\bz}\bZ$ by the horizontal parallelism; the integrability of the horizontal distribution is given by the following set of integrability constraints
\beq
 \cB_{1t}= i\epsilon (\cP_{2}\beta^{*}_{2}+\left(2\cP_{1}+\cK_{2}\right)\beta^{*}_{1}+\beta^{*}_{1x})
\,,\quad
\cB_{2t} =i\epsilon ( \cK_{1}\beta^{*}_{1}+\left(\cP_{1}+2\cK_{2}\right) \beta^{*}_{2} + \beta^{*}_{2x})\,, 
\\
\cA_{kt}=-i( \beta_{k}\left(\cP_{1}+\cK_{2}\right)
+\beta_{1}\cP_{k}+\beta_{2}\cK_{k}-\beta_{kx})\,,\quad
\cA_{k} = -\beta_{kt},\quad
\cB_{k} =\epsilon\beta ^{*}_{kt}\,, 
\\
\cC_{k} = -3\beta_{k},\quad
\cD_{k} = -3\epsilon\beta^{*}_{k}\,, \quad
\cP_{k} = i\epsilon \beta^{*}_{1}\beta_{k} + \eta_{k}\,, \quad
\cK_{k} = i\epsilon \beta^{*}_{2}\beta_{k} + \mu_{k}\,,
\eeq
where $\eta_{k}$ and  $\mu_{k}$ are arbitrary constants and $k=1,2$.
\ePf

Such constraints can be recasted in the form of families (depending on a matrix parameter) of coupled nonlinear Schroedinger equations:

\noindent 
$i\bbet_{x}+\bbet_{tt}+ \mathfrak{K}
\bbet +\epsilon |\bbet|^{2} \bbet = 0$,
$-i \bbet^{*}_{x} + \bbet^{*}_{tt} + \mathfrak{K}^{T} \,
\bbet^{*} + \epsilon |\bbet|^{2} \bbet^{*} = 0$,

\noindent
where $\bbet $ $=$ $( \beta_{1},  \beta_{2})^{T}$, $\bbet^{*}  $ $=$ $(\beta_{1}^{*},  \beta_{2}^{*})^{T}$, $ |\bbet|^{2} $ $=$ $ \bbet\cdot
\bbet^{*}$; 
the dot means the scalar product, subscripts stand for partial derivatives and $\mathfrak{K}$ is the matrix 
\beq
\mathfrak{K}=\left(
\begin{array}{ll}
\mathfrak{m}_{1} &  \kappa \\
 \kappa^{*} & \mathfrak{m}_{2}
\end{array}\right)\,.
\eeq

We require $\mathfrak{K}^{T}\,=\mathfrak{K}^{*}$: under this conditions the entries of 
the matrix $\mathfrak{K}$ have to be
all real, with $\mathfrak{m}_{1} = - i \left(2\eta_{1}+\mu_{2}\right)$ 
$\in \R$, 
$\mathfrak{m}_{2} = - i \left(\eta_{1}+2\mu_{2}\right)$ 
$\in \R$. 
In particular, 
$\eta_{2}=-\mu_{1}= -i\kappa$ and $\eta_{1}=-\mu_{2} =i
\mathfrak{m}$, with $\kappa, \mathfrak{m}$ real numbers, and $ \mathfrak{m}_{1} = - \mathfrak{m}_{2}$.

A representation of $\mathfrak{g}$ in fiber local coordinates on $\bP$ is given by
\\
$\T_{lm}^{(h)} =- i \sum^{+\infty}_{k=-\infty}
\xi^{(h+k)}_{l} \frac{\partial}{\partial \xi_{m}^{(k)}}+
i \del_{lm} \sum^{+\infty}_{k=-\infty}
\xi^{(h+k)}_{0} \frac{\partial}{\partial \xi_{0}^{(k)}}$, \\
with $h \in \Z$ and $\bxi^{(k)}$ $=$ $(\xi_{0}^{(k)}, \xi_{1}^{(k)}, \xi_{2}^{(k)})^{\rm T}$. 
We thus obtain equations for the absolute parallelism in the form of the linear spectral problem the integrability condition of which is the family of coupled nonlinear Schroedinger  equations we have derived:
$\bPsi_{x}(\lambda)= \cL_{1}\bPsi(\lambda)$, $\bPsi_{t}(\lambda)= \cL_{2}\bPsi(\lambda)$
where $\bPsi$ is the $3$-component vector  $\bPsi(x, t; \lambda) = 
\sum^{+\infty}_{k=-\infty}\,\lambda^{k}\bxi^{(k)}$ and with
{\footnotesize \beq 
&\cL_{1} &= \left(
\begin{array}{lll}
-i(\frac{1}{2}\epsilon |\beta_{1}|^{2}+\frac{1}{2}\epsilon|\beta_{2}|^{2}+ 6\lambda^{2}) &  
\beta_{1t}^{*}+ 3\lambda \beta_{1}^{*} &  \beta_{2t}^{*} + 3\lambda \beta_{2}^{*}\\
\frac{1}{2}\epsilon(- \beta_{1t} + 3\lambda \beta_{1}) & i\,(\frac{1}{2}\epsilon |\beta_{1}|^{2} 
+ 3 \lambda^{2}+\mathfrak{m_1}) &  i\,(\frac{1}{2}\epsilon \beta_{2}^{*}\beta_{1} + \kappa)\\
\frac{1}{2}\epsilon (- \beta_{2t} + 3\lambda \beta_{2}) &  i(\frac{1}{2}\epsilon \beta_{2} \beta_{1}^{*}\,
+ \kappa) & i(\frac{1}{2}\epsilon |\beta_{2}|^{2} + 3\lambda^{2} + \mathfrak{m_2})\\ 
\end{array}\right),
\\
&\cL_{2}&=\left(
\begin{array}{lll}
2\lambda &  i \beta_{1}^{*} &  i \beta_{2}^{*}\\
 \frac{ i}{2} \epsilon \beta_{1}  & -\lambda  &  0 \\
 \frac{ i}{2} \epsilon \beta_{2}  &  0 & -\lambda \\
\end{array}\right).
\eeq}

\noindent In particular, in the case $\mathfrak{m}_{1} = - \mathfrak{m}_{2} = \mathfrak{m}$ we obtain the spectral linear problem associated with the coupled nonlinear Schroedinger  equations 
describing the waves propagation in {\em twisted birefringent optical fibres} \cite{TriWWS89}
$\bPsi_{x}(\lambda)$ $=$ $ \cM_{1}\bPsi(\lambda)$, $\bPsi_{t}(\lambda)$ $=$ $ \cM_{2}\bPsi(\lambda)$, with $ \cM_{2}= \cL_{2}$ and
{\footnotesize \beq 
&\cM_{1} &= \left(
\begin{array}{lll}
-i(\frac{1}{2}\epsilon |\beta_{1}|^{2}+\frac{1}{2}\epsilon|\beta_{2}|^{2}+ 6\lambda^{2}) &  
\beta_{1t}^{*}+ 3\lambda \beta_{1}^{*} &  \beta_{2t}^{*} + 3\lambda \beta_{2}^{*}\\
\frac{1}{2}\epsilon(- \beta_{1t} + 3\lambda \beta_{1}) & i\,(\frac{1}{2}\epsilon |\beta_{1}|^{2} 
+ 3 \lambda^{2}+\mathfrak{m}) &  i\,(\frac{1}{2}\epsilon \beta_{2}^{*}\beta_{1} + \kappa)\\
\frac{1}{2}\epsilon (- \beta_{2t} + 3\lambda \beta_{2}) &  i(\frac{1}{2}\epsilon \beta_{2} \beta_{1}^{*}\,
+ \kappa) & i(\frac{1}{2}\epsilon |\beta_{2}|^{2} + 3\lambda^{2} - \mathfrak{m})\\ 
\end{array}\right).
\eeq}

\noindent We also notice that for  $\mathfrak{m}_{1}=-\mathfrak{m}_{2} =0$ we  recover, as a special case, the spectral problem 
$\bPsi_{x}(\lambda)=\cN_1 \bPsi(\lambda)$, $ \bPsi_{t}(\lambda)=\cN_2 \bPsi(\lambda)$, with $ \cN_{2}= \cL_{2}$
and

{\footnotesize \beq
&\cN_1&=  \left(
\begin{array}{lll}
-i(\frac{1}{2}\epsilon |\beta_{1}|^{2}+\frac{1}{2}\epsilon|\beta_{2}|^{2}+ 6\lambda^{2}) &  
\beta_{1t}^{*}+ 3\lambda \beta_{1}^{*} &  \beta_{2t}^{*} + 3\lambda \beta_{2}^{*}\\
\frac{1}{2}\epsilon(- \beta_{1t} + 3\lambda \beta_{1}) & i\,(\frac{1}{2}\epsilon |\beta_{1}|^{2} 
+ 3 \lambda^{2}) &  i\,(\frac{1}{2}\epsilon \beta_{2}^{*}\beta_{1} + \kappa)\\
\frac{1}{2}\epsilon (- \beta_{2t} + 3\lambda \beta_{2}) &  i(\frac{1}{2}\epsilon \beta_{2} \beta_{1}^{*}\,
+ \kappa) & i(\frac{1}{2}\epsilon |\beta_{2}|^{2} + 3\lambda^{2} )\\ 
\end{array}\right), 
\eeq} 

\noindent the compatibility condition of which produces a family of coupled nonlinear Schroedinger  equations describing \eg  waves propagations in birefringent nonlinear optical fibers \cite{TriWWS88,TriWWS89}.

It is noteworthy that for $\mathfrak{K}=0$ we obtain as an element of the family the well known {\em Manakov equation} ({\em and} the associated spectral linear problem) which then {\em shares with coupled nonlinear Schroedinger equations  the same tower with skeleton}. This fact is much stronger than simply knowing that solutions of the coupled nonlinear Schroedinger equations could be reconstructed (resorting \eg to Miura-type or gauge transformations) by solutions of the Manakov equation (a local fact; while our result is of intrinsic global and deeper nature). 

\bRm
We point out that the seeking for the representation $\rho$ of a skeleton belonging to the tower $\bP(\bZ,\bG)$ generates a whole family of equations of physical significance and also associates it explicitly with a linear spectral problem. We also notice that, consequently, different {\em ans\"atze}  on 
$\bF$ and $\bG$ should be interpreted as the seeking for different representations; the latter would provide possibly different families of equations.
\eRm

Here we provide a geometric interpretation in terms of the construction of a tower 
with skeleton and an associated gauge-natural bundle of the {\em ansatz}  introduced for the first time in \cite{Pa93} and used in \cite{ALLPS95} .  We also stress that dispersive nonlinear equations have been obtained as necessary conditions for our original algebraic structure to be a skeleton for an integrable tower. We remark that the request that the operator $\T (\cdot)$ satisfy the Ka\v{c}-Moody algebra can be weakened to the request that they satisfy the `open' Lie algebra structure, obtaining, nevertheless, dispersive nonlinear equations as integrability conditions for a tower.

This geometric interpretation of the so-called `inverse prolongation procedure' could be used in order to couple in a non trivial way different fields: one would ask whether `open' Lie algebra structures obtained from the prolongation of two  different field equations could be provided of the structure of a {\em unique} skeleton for a tower the integrability condition of which would provide {\em new} field equations. This topic is currently under investigation.

\end{document}